\documentclass[twocolumn]{aa}
\pdfoutput=1
\usepackage{graphicx}
\usepackage{txfonts}
\usepackage{natbib}
\newcommand{\dk}{}

\newcommand{\resub}{\bf }
\begin{document}

\title{Is the solar spectrum latitude dependent?}
\subtitle{An investigation with SST/TRIPPEL}
\author{ 
    Dan Kiselman \inst{1,2} 
\and    Tiago M. D. Pereira  \inst{3}\thanks{Current Address: NASA ARC, Mail Stop 245-3, P.O. Box 1, Moffett Field, CA 94035, USA}
\and    Bengt Gustafsson \inst{4,5}
\and    Martin Asplund \inst{6,3}
\and    Jorge Mel{\'e}ndez  \inst{7}
\and    Kai Langhans \inst{1,2}\thanks{Current address: Gothmunder Weg 8, 23568 L{\"u}beck, Germany} 
 }

\institute{
                The Institute for Solar Physics of the Royal Swedish Academy of Sciences, AlbaNova University Centre, SE-106~91 Stockholm, Sweden \and
                Stockholm Observatory, Department of Astronomy, Stockholm University, AlbaNova University Centre, SE-106~91 Stockholm, Sweden \and
                 Research School of Astronomy and Astrophysics, Australian National University, Cotter Rd., Weston, ACT 2611,   
                 Australia \and
                Department of Physics and Astronomy, Uppsala University, Box 515, SE-751\,20 Uppsala, Sweden \and
                Nordita, AlbaNova University Centre, SE-106~91 Stockholm, Sweden \and
                Max-Planck-Institut f\"{u}r Astrophysik, Postfach 1317, 85741, Garching b. M\"{u}nchen, Germany \and                   
Departamento de Astronomia do IAG/USP, Universidade de S\~ao Paulo,
  Rua Mat\~ao 1226, S\~ao Paulo, 05508-900, SP, Brazil
}

\date{Received  / Accepted }
\offprints{Dan Kiselman,
\email{dan@astro.su.se}}

\authorrunning{Kiselman et al.}
\titlerunning{Is the solar spectrum latitude dependent?}

\abstract
{ In studies of the solar spectrum relative to spectra of solar twin
  stars, it has been found that the chemical composition of the
  Sun seems to depart systematically from those of the twins. One
  possible explanation could be that the effect is due to the special
  aspect angle of the Sun when observed from Earth, as compared with
  the aspect angles of the twins.  Thus, a latitude dependence of the
  solar spectrum, even with the heliocentric angle constant, could lead to effects of the
  type observed.  }
{ We explore a possible variation in the strength of certain spectral
  lines, used in the comparisons between the composition of the Sun
  and the twins, at loci on the solar disk with different latitudes
  but at constant heliocentric angle.  }
{ We use the TRIPPEL spectrograph at the Swedish 1-m Solar Telescope
  on La Palma to record spectra in five spectral regions in order to
  compare different locations on the solar disk at a
  heliocentric angle of 45\degr. 
  Equivalent widths and other
  parameters are measured for fifteen different lines representing nine
  atomic species. Spectra acquired at different times are used in
  averaging the line parameters for each line and observing position.
}
{
 The relative variations in equivalent widths 
at the equator and at solar latitude $\sim 45\degr$ are found to be less  than $1.5\%$  for all
spectral lines studied. Translated to elemental abundances as they
would be measured from a terrestrial and a hypothetical pole-on
observer, the difference is estimated to be within 0.005 dex in all cases.}
{It is very unlikely that latitude effects could cause the reported abundance difference between the Sun and the solar twins.  The accuracy obtainable in measurements of  small differences in spectral line strengths between different solar disk positions  is very high, and can be exploited in studies, e.g. of weak magnetic fields or
effects of solar activity on atmospheric structure.}

\keywords{Sun: abundances -- 
          Sun: atmosphere -- Sun: spectrum --
          techniques: spectroscopic 
}

\maketitle


\section{Introduction}
\label{sec:intro}
In recent studies it has been found that the chemical composition of
the Sun, as compared with those of very similar stars, so-called solar
twins, is systematically different: the Sun is comparatively less rich
in refractories, i.e. elements that easily condensed to solids for
instance in the solar protoplanetary nebula, than in volatiles
\citep{2009ApJ...704L..66M,2009A&A...508L..17R}.  The range of the
effect is about $20\%$ for volatiles as compared with refractories.
Several possible explanations for this interesting phenomenon have
been discussed \citep[see][]{2009ApJ...704L..66M,2010Ap&SS.tmp...36G},
including cleansing
of the protoplanetary disk by planetesimal formation. 
A more mundane explanation could be an
obvious selection effect: in Earth-based studies the Sun is by
necessity observed from a position close to the solar equatorial
plane. The solar twins, on the other hand, are presumably observed
with aspect angles randomly distributed with respect to their rotation
axes. Thus, if the solar spectrum, integrated across the disk, for
some reason were somewhat different as seen from a position far from
 the equatorial plane, the differences interpreted as abundance
differences might have this explanation instead. An empirical way of
exploring this is to observe solar photospheric spectra at different
disk positions, though with a common angular distance from the disk
centre so that the first-order centre-to-limb variations can be
directly compensated for. Here, such a study is presented.

Studies of centre-to-limb variation of the solar spectrum have
been pursued for more than a century
\citep[cf.][]{1873Natur...8...77H, 1907ApJ....25..300H}
Some observational studies have been devoted to the question whether there are any
differences between spectra from regions along the solar equator as
compared with along a meridian.  These studies have mostly been
concentrated towards variations in line shifts and line shapes
(bisectors) \citep[e.g.][]{1976SoPh...46...29C,1978SoPh...57...13C,
  1980SoPh...68...41B,1982SoPh...79....3B,1984SoPh...94...49A},
variations that {\dk have generally been} interpreted as variations in the
photospheric velocity fields with latitude.
There are few studies of the variations, or constancy, in the line
strengths (equivalent widths).  One notable example is the study by
\citet{1994A&A...283..263R} of the variation along the equator and a
meridian of four lines of \ion{C}{i}, \ion{Si}{i}, \ion{Mn}{i} and \ion{Fe}{i}.  In addition to
studies of line widths and bisectors, these authors also included
measurements of the variations of the equivalent widths. For \ion{Si}{i}, \ion{Mn}{i} 
and \ion{Fe}{i} lines they found the lines to be a few percent weaker along the
equator than along the meridian at given $\mu\approx 0.6$, while the \ion{C}{i}
 line showed less variation. Although not very pronounced this
tendency has the direction needed to explain the apparently low ratio
of refractories/volatiles in the Sun compared to the twins. However,
three of the four spectral lines studied by Rodr\'iguez Hidalgo et
al. are on the flat part of the curve of growth, while the lines
used in the analyses of \citet{2009ApJ...704L..66M} and 
\citet{2009A&A...508L..17R} are almost all on the linear part with line
strengths insensitive to the velocity fields (among the lines of Mel\'endez et al. is \ion{C}{i}
538.0322~nm which is the fourth line measured by Rodr\'iguez Hidalgo et
al.).  Therefore, the qualitative tendency found by Rodr\'iguez
Hidalgo et al.  cannot be used to infer what could be the effects on
abundances derived from solar flux spectra at different aspect
angles. Instead, a direct comparison of line strengths of the mostly
weak lines used in the Sun--twin abundance comparison is needed.

It is clear from the previous work on the latitude dependence of the
solar spectrum that considerable care must be exercised in positioning
the spectrometer slit and averaging properly over granulation and
other inhomogeneities. A strictly differential approach must be
adopted, in order to minimise the effects of stray light and other
instrumental issues.

\section{Observations}
\label{sec:obs}
The observations were made with the
Swedish 1-m Solar Telescope (SST) on La Palma \citep{scharmer03new}
employing the TRIPPEL spectrograph during the period of 3--15 May 2010. 
Since this instrument has not yet been subject to
a detailed description in the literature, we use this opportunity to give a summary of its 
properties.

\subsection{The TRIPPEL spectrograph}
The TRI-Port Polarimetric Echelle-Littrow (TRIPPEL) spectrograph
had the following design goals:
\begin{itemize}
\item allow simultaneous observations at three separate wavelengths
\item in principle exploit the full spatial resolution of the SST
\item wavelength range about 380\,nm -- 1100\,nm
\item good polarimetric properties
\item moderate (for a solar spectrometer) spectral resolution ($R \approx 200,000$)
\end{itemize}

\begin{figure}
   \centering
   \includegraphics[width=\columnwidth]{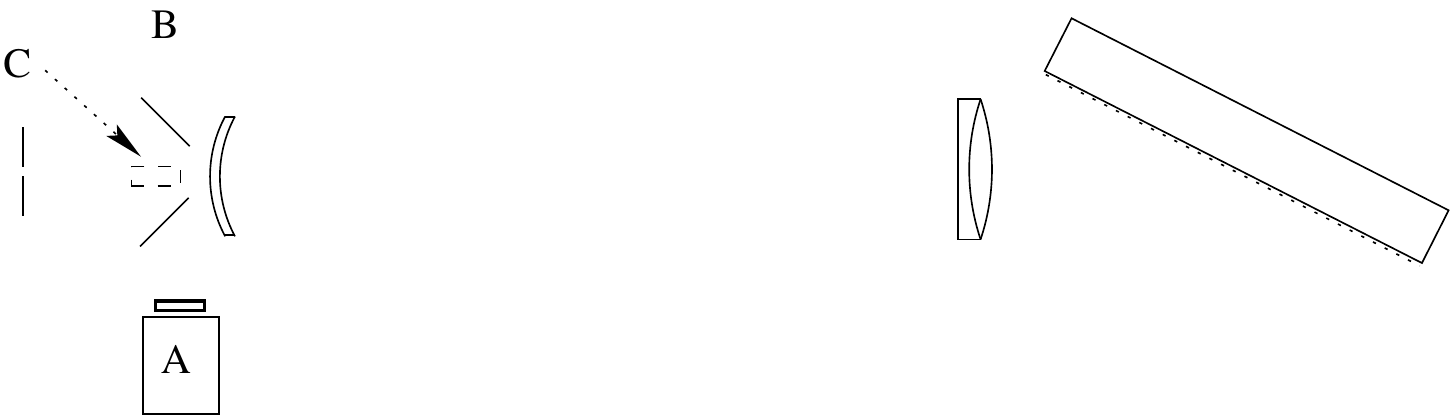}
   \caption{Schematic view of the TRIPPEL spectrograph as seen from
     above. Components from left to right: entrance slit, pick-off mirrors for
   ports A-C, field lens, Littrow doublet, grating. Spectral port A has a CCD camera with an interference
     filter. }
              \label{fig:trippelsketch}%

\end{figure}

To achieve this, it is designed as a lens Littrow spectrograph
with a ruled echelle grating. The design draws on the experience from 
the short Littrow spectrograph at the SVST \citep[the predecessor of
the SST]{scharmer85concepts}. Some relevant numbers are given in
Table~\ref{table:trippel} and a schematic view of the main optical
designis shown in Fig.~\ref{fig:trippelsketch}.

The slit is etched on a chromium-coated glass plate that is
easily interchangeable and the reimaging lens of the SST could also in
principle be changed to give another image scale.  The spectrograph
optics and the grating can accommodate a beam that is twice as
fast as was used in the current setup (f/46). The slit plate can be
rotated around a vertical axis to allow the reflected light to be
directed towards slit-jaw imagers. It can also be rotated around an
horizontal axis orthogonal to the optical axis to adjust the tilt of
the slit relative to the dispersion direction. There are two movable
jaws behind the slit plate that can be used to cover parts of the slit
to minimise the amount of light entering the spectrograph, and to allow
dark areas on the spectrum camera for reference.

The Littrow doublet lens is mounted on a translational stage. To get
the required large diffraction-limited field-of-view, a meniscus field
lens is placed close to the focal plane so that the light passes twice through
it. The optical design fixes the Littrow lens position relative to
this field lens. The translational stage is thus only used when
checking the spectrograph focus and never for focusing the spectrograph
cameras.

The grating from Richardson Grating Laboratory has
excellent polarization properties, with the
maximum difference between S and P efficiency being about 10\,\% in
the visual region. The grating is mounted on a rotational stage with
0.01\degr\ setting precision.

Three silver-coated mirrors pick off light to the exit ports.  
For the two ports with horizontal light beams
(designated A and B), the cameras are mounted on holders sliding
on a rail of the same
type as is used for the SST imaging setups, thus allowing freedom in
camera choice.  The third port (C), which has a vertical light beam,
currently uses a special holder for a Redlake
Megaplus\,{\sc ii} camera (ES1603).

Apart from some image distortion caused by the corrector lens,
TRIPPEL has an off-plane design which leads to curvature of
the spectral lines (smile) and also of the dispersion direction
(keystone). The curvature over the typical CCD size is small,
so the greatest effect of the smile is to cause the spectral lines to
tilt relative to the axis defined by the direction of the grating grooves. 
The slit is rotated about 
3.5\degr\ from the vertical direction to partly compensate for this.

Since all spectral orders fall on top of each other in the focal plane, interference
filters placed just in front of the CCDs are used to select the
desired order. This also has the advantage of blocking straylight outside
the filter passband. The transmission profile of two-cavity filters typically has a base that
is too wide to be practical, so three-cavity filters are used.
Assuming a profile for a
three-cavity interference filter, requiring that the transmission
at a distance of a full spectral range from the central wavelength is
$10^{-4}$ of its peak value, and applying a significant margin of
safety, result in the filter widths in
Table~\ref{table:trippelparams}.

\begin{table}

\caption{Some properties of the TRIPPEL spectrograph}             
\label{table:trippel}      
\centering                          
\begin{tabular}{r l}        
\hline                        
 Slit width & 25 $\mu$m ($\approx$ 0\farcs11) \\
 Grating constant & 79.01 mm$^{-1}$\\
 Blaze angle & 63.43\degr \\
 Grating size & $154\times 306$ mm\\
 Focal length & 1500 mm \\
 Littrow lens aperture & 135 mm \\
\hline                                   
\end{tabular}
\end{table}

\begin{table}
\caption{Nominal TRIPPEL parameters at the wavelengths of some interesting
  lines.}             
\label{table:trippelparams}      
\centering                          

\begin{tabular}{lccrccc}        
\hline\hline                 

{\resub Line} & $\lambda$  & {\sc fwhm}   & $n$ & Bandpass & $dx/d\lambda$ & $W$ \\ 
     & [nm] & [nm] & & [km/s] & [mm/nm] & [nm] \\
\hline                        
   \ion{Ca}{ii} & 393.39 & 2~~ &  57&  1.3& 14.9& 0.93\\
   \ion{Mg}{i}   & 457.11 &  2.5 &  49&  1.3& 13.0& 1.06\\
   \ion{Mg}{i}   & 517.27 & 3~~ &  44&  1.2& 11.9& 1.16\\
   \ion{Fe}{i}   & 630.20 & 5~~ &   36&  1.2&  ~9.6& 1.44\\
   \ion{Li}{i}   & 670.78 & 6~~ &  34&  1.2&  ~9.3& 1.49\\
   \ion{Ca}{ii}  & 866.20 & 10~~ &  26&  1.3&  ~6.7& 2.05\\ 
\hline                                   
\end{tabular}
\tablefoot{ {\sc fwhm}: recommended width of a 3-cavity interference
  filter. $n$: spectral order. Bandpass: slit-limited. $dx/d\lambda$:
  dispersion. $W$: wavelength coverage for a 13.8 mm CCD (Megaplus 1.6).}
\end{table}

The dispersion decreases with decreasing order and thus with
increasing wavelength. The theoretical spectral resolution is limited
by the slit width and so stays rather constant between different
orders: it is between 1.1 and 1.4 km/s or $R={\lambda \over
  \delta\lambda} \sim 220000$. The resolution as found by our
straylight-fitting routine is generally lower than this, but there is
probably cross-talk between the spectral resolution and the straylight
parameter in our model.

The spectral interval recorded by each camera is determined by the turning
angle of the grating and the filter. With three ports, many combinations
are possible, especially because some freedom of camera movement in the dispersion direction is
available.

Table~\ref{table:trippelparams} lists some nominal {\resub instrumental} parameters at the
wavelengths of some interesting lines.

The spectrograph is focused by pressing a perpendicular slit against the
spectrograph slit. This creates a small
rectangle that is seen on the spectrum cameras as a very narrow
spectrum. The Littrow lens is then moved between two out-of-focus
position for a range of camera positions. The camera position where
the two lens positions create spectra of similar sharpness is taken as
its focus position. With some training it is not difficult to assess
the sharpness from the spectrum image displayed on the screen.

The other crucial focusing, that of the solar image on the 
spectrograph slit, is done with a similar technique. But instead of
placing a physical slit crossing the spectrograph slit, an {\em image}
of a slit is placed there by mounting a physical slit in the telescope focus at the exit port of the vacuum
system.
The adaptive optics is locked on a pinhole next to that slit.
The wavefront sensor is then adjusted (which leads to the image moving back and forth)
until symmetry in the out-of-focus spectra is achieved. 

Finally, TRIPPEL allows some auxiliary instruments.
The wavefront sensor that controls the adaptive mirror and the
correlation tracker camera are fed via a beamsplitter cube in front of
the slit.
The reflective slit plate makes slit-jaw imaging possible. For the current
observations, we employed up to two slit-jaw cameras fed with
reimaging optics. These cameras were equipped with interference
filters with central wavelengths close to the spectral regions
observed. The slit-jaw imaging was used for pointing control, and for
keeping track of the seeing quality and the light-intensity
level.

%
%
%
%
%
%

\subsection{Spectral lines}

\begin{table}
\caption{Spectral lines observed }             
\label{table:lines}      
\centering                          
\begin{tabular}{r l r r l}        
\hline\hline                 
{\resub Spectral} & {\resub Species} & CWL & $E_l$ & {\resub Class}\\    
region & & [nm] & [eV] \\
\hline                        
538 & \ion{C}{i} & 538.03 & 7.58 & volatile \\      
616 & \ion{Na}{i} & 616.08 & 2.10 & intermediate\\
616 & \ion{Fe}{i} & 616.54 & 4.14 & refractory \\
616 & \ion{Ca}{i} & 616.64 & 2.52 & refractory \\
616 & \ion{Ca}{i} & 616.90 & 2.52 & refractory\\
711 & \ion{C}{i} & 711.14 & 8.64 & volatile \\
711 & \ion{C}{i} & 711.32 & 8.65 & volatile \\
768 & \ion{S}{i} & 768.61 & 7.87 & intermediate \\
768 & \ion{K}{i} & 769.90 & 0.00 & intermediate \\
777 & \ion{O}{i} & 777.19 & 9.15 & volatile\\
777 & \ion{O}{i} & 777.41 & 9.15 & volatile\\
777 & \ion{O}{i} & 777.53 & 9,15 & volatile\\
783 & \ion{Al}{i} & 783.53 & 4.02 & refractory \\
783 & \ion{Al}{i} & 783.61 & 4.02 & refractory \\
869 & \ion{S}{i} & 869.46 & 7.87 & intermediate\\
\hline                                   
\end{tabular}
\end{table}

\begin{table*}
\caption{Observational setups and the number of data sets from each
  day used in the
  final analysis.} 
\label{table:obs}      
\centering                          
\begin{tabular}{r r r r c c c c c c c c c c}        
\hline\hline                 
 & Port A & Port C & Port B &DC & N & NW& W & SW& S & SE& E & NE&  \\    
Date &  &  &             &$-3\degr$
& 42\degr & 27\degr&-3\degr&$-32\degr$&$-48\degr$&$-32\degr$&$-3\degr$&$27\degr$ \\
\hline                        
2010-05-05 & 538 & 711 & 783    & 5 & 1 & 1 & 1 & 1 & 1 & 1 & 0 & 1 &  \\
2010-05-06 & 538 & 711 & 783    & 9 & 3 & 3 & 3 & 3 & 3 & 3 & 3 & 2 &  \\
2010-05-07 & 538 & 711 & 783    & 7 & 3 & 1 & 3 & 1 & 3 & 1 & 3 & 1 &  \\
2010-05-09 & 768 & {\resub ...} & 777    & 6 & 5 & 0 & 4 & 0 & 5 & 0 & 5 & 0 &  \\
2010-05-14 & 768 & {\resub ...} & 777    & 1 & 1 & 0 & 0 & 0 & 0 & 0 & 0 & 0 &  \\
2010-05-15 & 869 & {\resub ...} & 616    & 7 & 4 & 0 & 4 & 0 & 4 & 0 & 4 & 0 &  \\
\hline                                   
\end{tabular}
\tablefoot{DC is solar disk centre. Solar
  latitudes are given for each disk position.}
\end{table*}

In the present study we have selected some spectral lines 
used by  \citet{2009ApJ...704L..66M}
in their analysis of the Sun and solar twins, ensuring that both refractories,
volatiles, and intermediate elements were represented and that as
many lines as possible were observable simultaneously. The spectral regions were
ranked in priority and the setup was changed to the next set of regions when
enough data had been secured. Table~\ref{table:lines} lists the lines
observed and Table~\ref{table:obs} the setup combinations used.
Fig.~\ref{fig:EWregions} shows plots of all the lines in their spectral context of
our solar observations.

   \begin{figure}
   \centering
   \includegraphics[width=\columnwidth,bb= 0 0 288 466]{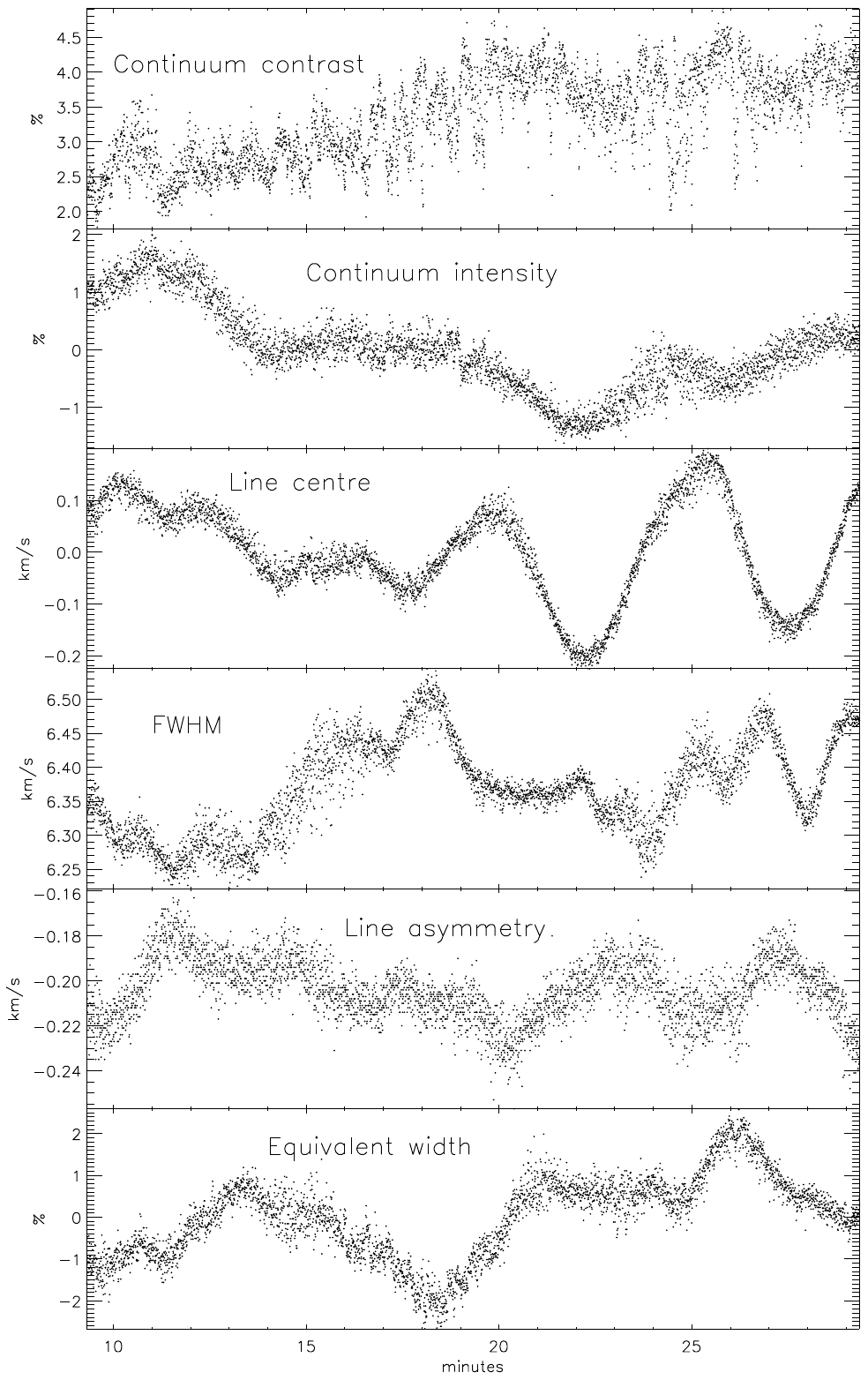}
   \caption{{\resub Temporal} sequence of the \ion{Al}{i} 783 nm line at disk center. The six quantities
     are plotted as function of time in minutes after 10:00~UT. 
     Equivalent width and continuum intensity are plotted as the
     deviation in percent from their respective mean values. Each
     point corresponds to the mean spectrum along the slit for one exposure.}
              \label{fig:long}%
    \end{figure}

\subsection{Observations: strategy and practice}
We assume that a latitude dependence of line equivalent widths that is strong enough to give
a significant effect in the flux spectrum, must be present also at a
moderate distance from solar disk centre. We choose a heliocentric
distance of 45 degrees or $\mu = \sqrt{2}/2 \sim 0.71$. For 
the solar meridian, this corresponds to the point that is seen
at the same $\mu$ for an observer in the ecliptic plane and
one that would observe the Sun pole-on. It is also close to $\mu = 2/3$,
which is where, according to the Eddington-Barbier relations, the outgoing
intensity spectrum should be characteristic of the flux spectrum 
and thus relevant for stellar observations.


In order to decrease bias from north-south and east-west systematic differences, perhaps
due to random activity patterns, we selected eight different position angles and
named these points N, W, S, E, NW, SW, SE, and NE. To test the stability of the results,
the solar disk centre was observed frequently and interspersed with the $\mu = 0.71$ pointings.
The solar latitudes of the observed points are given in Table~\ref{table:obs}.
The heliographic latitude at the time of observations was
approximately $-3\degr$.

An observational sequence consisted of a series of pointings with
about 5~min spent at each. Before and after each sequence flat fields were acquired.
The sequence typically started at disk centre and
then going to all the eight $\mu=0.71$ points, after which disk centre was
observed again.
All the time during observations, the Ca\,{\sc ii} H imaging camera
was used to make sure that the regions observed were quiet.

After the first days of observations when inspection of the data confirmed that
the stability of the results was quite high, the intermediate points (NW, SW, SE, NE) were abolished
from the observing sequence, and only observations in the N, W, S, and E positions, as well as the disk centre,
were made.

The first data sets were taken with fixed pointing. In order to
increase the spatial sampling, a procedure where the slit was scanning
back and forth over 3\arcsec was implemented. This was done via the
correlation tracker.

   \begin{figure*}
   \centering
   \includegraphics[bb= 0 0 576 576,width=\textwidth]{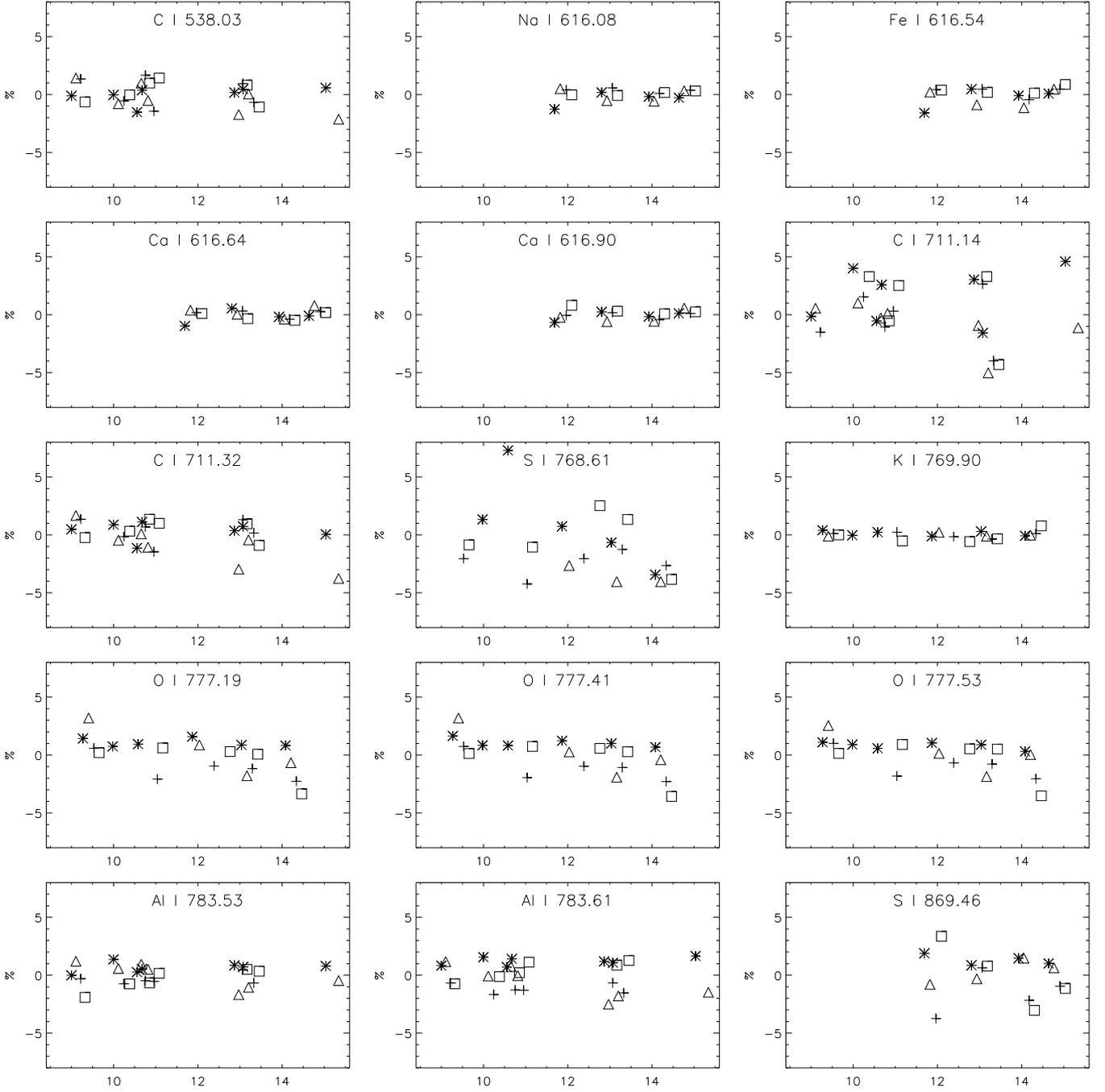}
   \caption{Equivalent widths for all lines and data sets for $\mu =
     \sqrt{2}/2$. 
      The deviation from the mean value from each data
     set is plotted in units of percent of the overall mean value and
     as a function of the time of day in hours 
     when it was acquired. Symbols according to solar disk position. Asterisk: N, Plus: S, Triangle: W, Square: E.}
              \label{fig:plotw}%
    \end{figure*}

\subsection{Data reductions}
The data reduction procedures follow those described by
\citet{2009A&A...507..417P} with few deviations.
Each data set requires a set of flat-field and dark exposures. The
flat-fields are acquired by taking many exposures while the telescope
is scanning over a designated area around disk centre
that should be devoid of spots.  For the current observations, a new procedure
was used where the adaptive mirror of the SST is fed with random
voltages in order to smear out spatial structures even more.
The darks are acquired with the light beam blocked close to the
telescope exit window.

The flat-field exposures are used to compute {\em gain tables} and
mapping the {\em geometric distortions} of the spectrograms.
They are also used for
wavelength calibration and scattered-light corrections in a procedure similar to that of
\cite{2004A&A...423.1109A}, thus comparing
the observed spectra with  
the Fourier Transform Spectrometer at the McMath-Pierce 
Telescope (hereafter called the FTS atlas) of \citet{fts-atlas}.

{\em Wavelength calibration} is done by fitting line cores. The
resulting wavelength scale is thus that of the FTS atlas.

{\em Straylight} inside the spectrograph is treated as a constant which is determined
simultaneously with the spectral resolution by a least-squares fit to
the reference spectrum. Typical values are 5-6~\% of the spectral continuum.

{\em Telluric lines} are identified and excluded from the fitting procedures.

Remaining artefacts in the continuum level -- caused for
example by residual components of fringes perpendicular to the dispersion direction-- are also removed
with the help of the reference spectrum. The result is a set of parameters
that maps the observed disk-centre average spectrum to that of the atlas.

\begin{figure*}
   \centering
   \includegraphics[width=\textwidth]{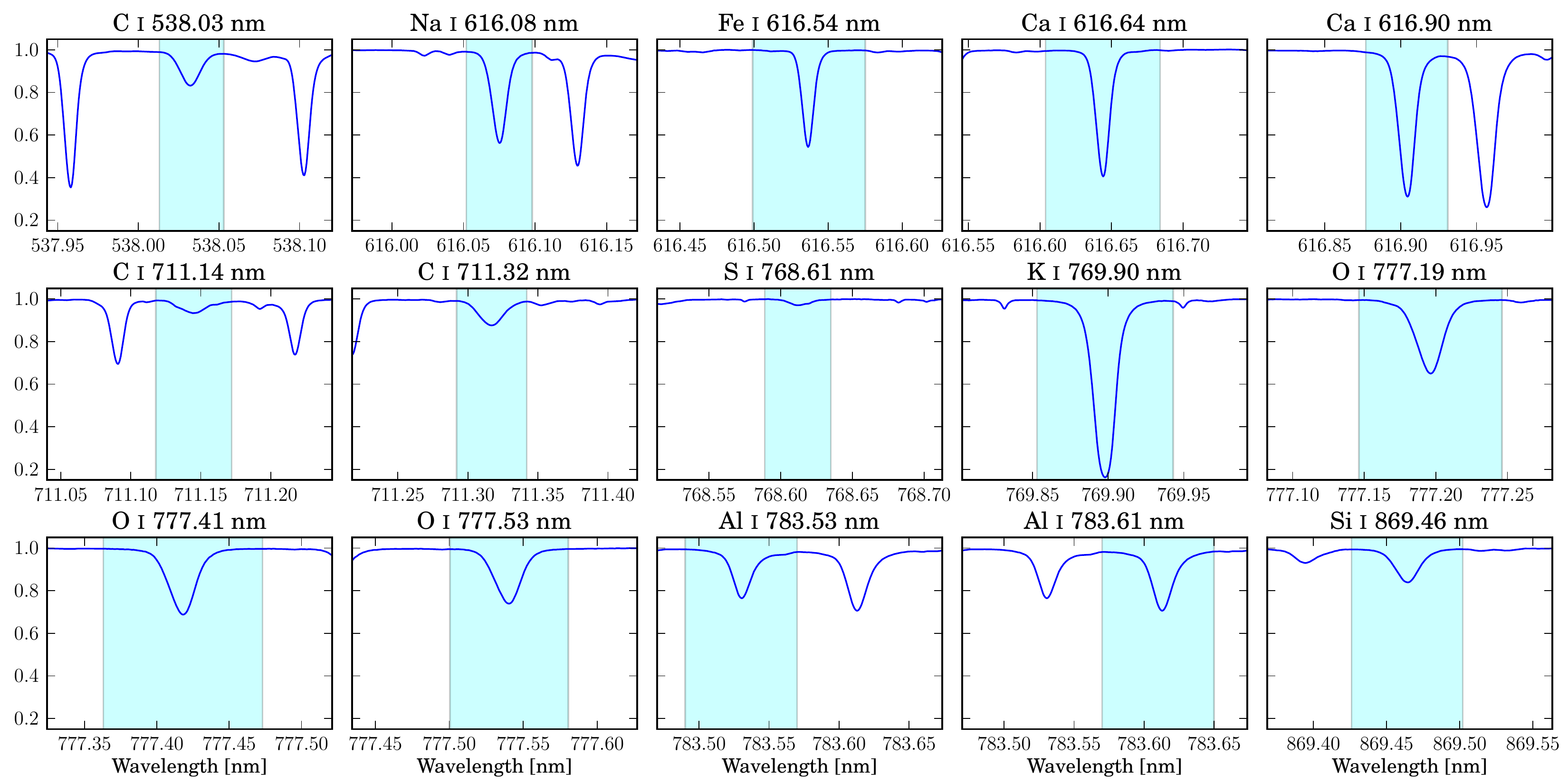}
   \caption{All spectral lines of the present study. Shaded region shows the
     integration range for the equivalent widths. This range is
     shifted according to the line core for every individual
     spectrum.}
              \label{fig:EWregions}%
    \end{figure*}

The process described above produces spectrograms ready for analysis.
In the current work, we proceed by coadding each spectrogram over
the spatial direction to get a one-dimensional spectrum that is the
average spectrum under the spectrograph slit. A number of these spectra are
then coadded, thus forming an average also over time.
Various quantities for the spectral lines of interest are then calculated
from such average spectra:
{\em Line centre} is the velocity difference between the local line core and the calibration spectra --it is thus relative to the FTS atlas;
{\em FWHM} is the usual full width at half maximum line depth;
{\em Line asymmetry} is the distance between the local line core and the line bisector
at half maximum line depth;
{\em Equivalent widths} are measured by direct integration over a predefined
interval which is centred on the line core and thus follows any local
Doppler shift -- this is illustrated in Fig.~\ref{fig:EWregions}.

\subsection{Data selection and error discussion}
\begin{figure}
   \centering
   \includegraphics[width=\columnwidth]{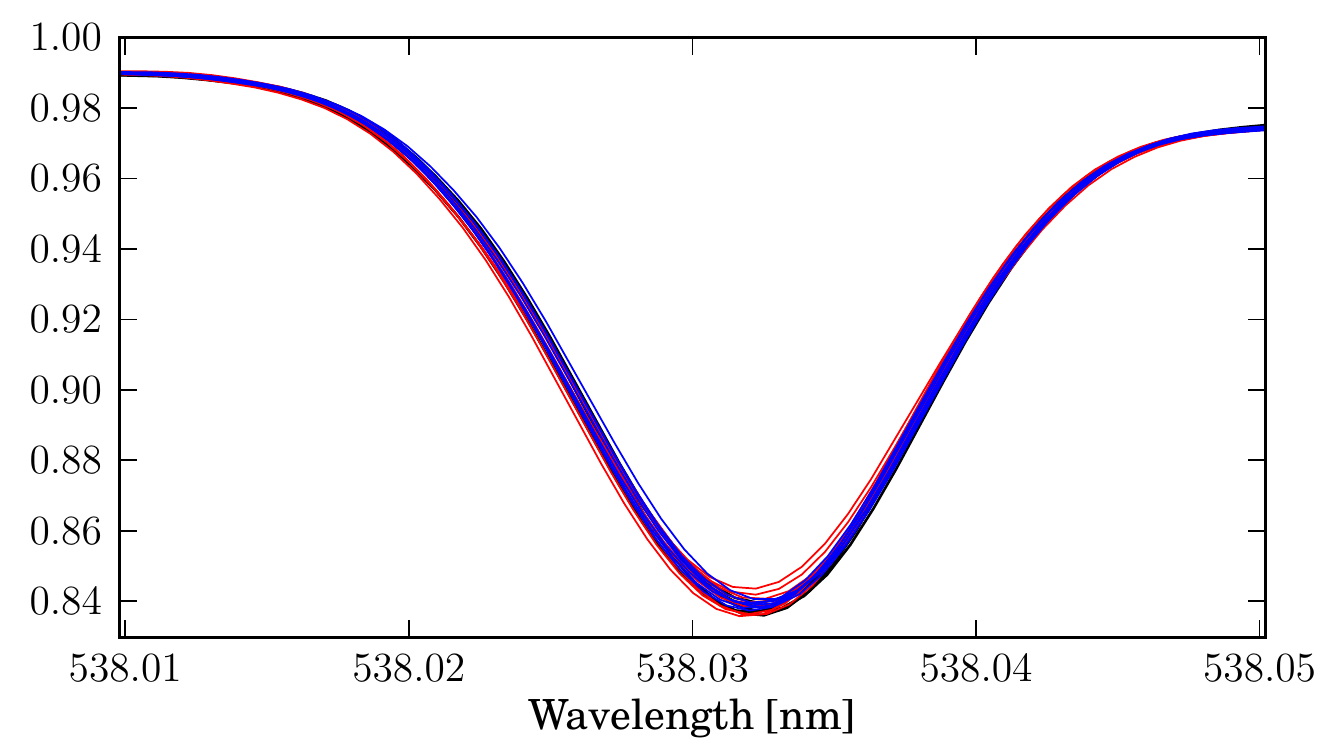}
   \caption{Disk-centre spectra for \ion{C}{i} 538 nm. Black
     2010-05-05, Blue 2010-05-06. Red 2010-05-07. }
              \label{fig:CI538}%
    \end{figure}

\begin{figure}
   \centering
   \includegraphics[width=\columnwidth]{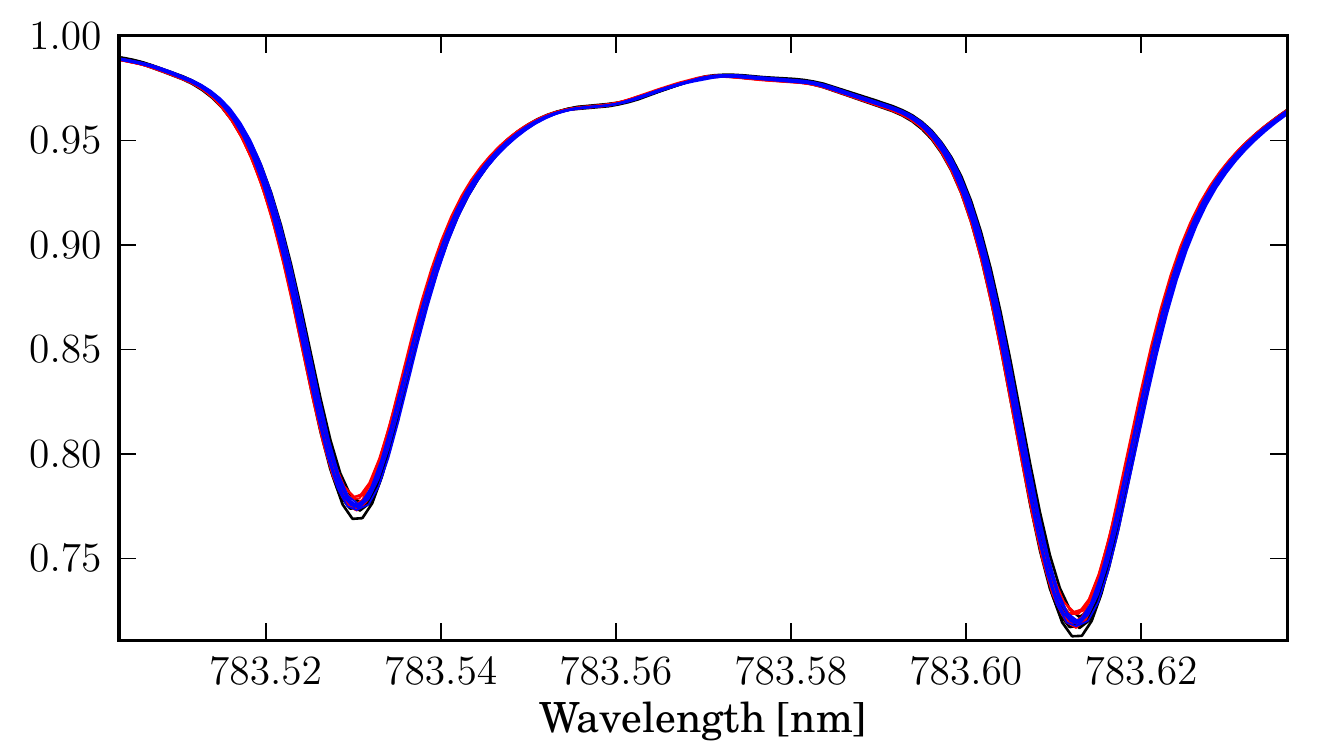}
   \caption{Like {\resub Fig.~\ref{fig:CI538}} but for the two lines of \ion{Al}{i} at 783 nm.  }
              \label{fig:AlI783}%
    \end{figure}

Sources of errors and other effects that would undermine the
observational strategy could be imprecise and non-reproduceable pointing,
seeing, variations on the Sun from granulation, oscillations, or
activity, and instrumental drifts. Figures~\ref{fig:CI538}--\ref{fig:AlI783}
illustrates this by showing several disk-centre spectral profiles of three lines.

\subsubsection{Pointing}
The exact pointing of the telescope is of utmost importance since the
equivalent width of the most sensitive lines in our sample -- the 
\ion{C}{i} lines at
711~nm -- is predicted to change by 5\% for a 1\arcmin shift ($\Delta \mu \approx 0.05$) along
the radius vector. 
We used the solar limb in four positions to calibrate the telescope
pointing before each sequence. After such a sequence we returned to
the solar limb to estimate the drift in pointing during the
sequence. Sometimes this error was found to be small but it could
typically amount to 25\arcsec.

The spectrograph slit is fixed in the laboratory frame and its orientation in the
sky plane will change during the day. No effort was made to measure
the orientation. The finite slit length ($\sim 50\arcsec$) thus adds to
the positional {\dk coverage of each spectrum, but does not affect the mean position of the slit.}

\subsubsection{Telluric lines}
\begin{figure}
   \centering
   \includegraphics[width=\columnwidth]{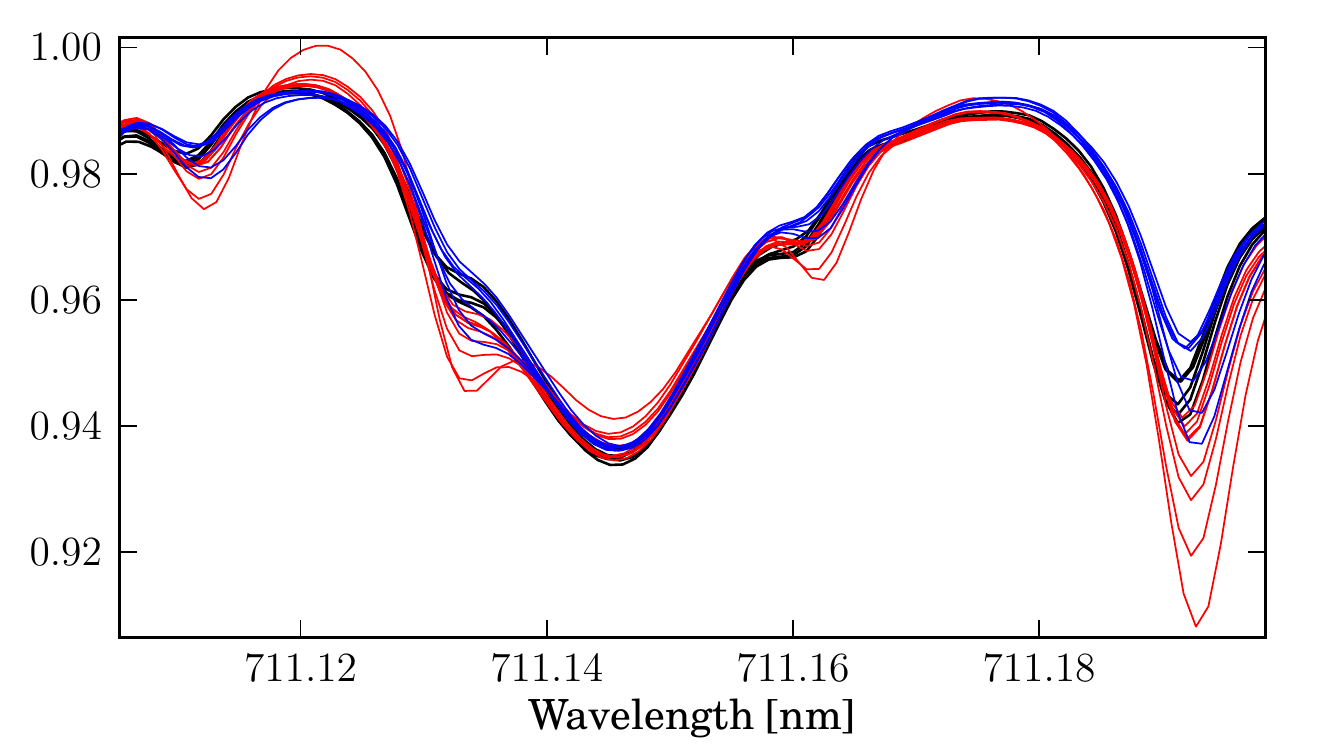}
   \caption{Like Fig.~\ref{fig:CI538} but for \ion{C}{i} 711.14 nm. Note the presence of telluric lines whose strengths varies with the air mass.}
              \label{fig:CI711}%
    \end{figure}

If a line is significantly blended with a telluric line, its
measured properties will be a function of the airmass and thus the
time of the observations. This is obvious for the \ion{C}{i}~711.14 nm
line as seen in Fig.~\ref{fig:CI711} {\dk (where the standard deviation of $\Delta W/W
= 2.8\%$)}
, but still this airmass dependence drowns in
the noise for this weak line since no clear trend with time is visible
in Fig.~\ref{fig:plotw}.

\subsubsection{Impact of seeing}
The ever-changing seeing causes the spatial resolution to vary with
time. Because the current project does not require very high spatial
resolution -- indeed some spatial averaging is necessary and beneficial -- one might
surmise that seeing is irrelevant. We did not take that approach. Bad
seeing implies that light from far away on the solar disk will be
scattered into the spectrograph slit without any control. We therefore
compared results from periods with different seeing
but could not discern any significant differences in the measured line properties.
Still
we excluded the data sets with the worst seeing, and for those that were
included, we kept the 50 best frames (100 for disk-centre spectra) as selected from continuum
contrast values. We note that really bad seeing can cause tracking
loss which itself is a source of error.

We caution, however, against assuming that seeing is unimportant in
studies like this. We
note that in our case the observed points were far from the limb, that
the centre-to-limb variation of the equivalent widths are rather linear (thus symmetric
blurring will still give the same results), and that the intensity
spectrum at this $\mu$ should anyway be similar to the flux spectrum.

A similar comparison of data from periods where thin high clouds were
intermittently in view gave a different result. Here the spectra did
change significantly and any cloud-affected observations were
subsequently excluded from analysis.

\subsubsection{Solar variations and instrumental stability}
Intrinsic variations of the solar target and instrumental drifts can in
principle be difficult to disentangle.  Figure~\ref{fig:long} shows
the result of a run where the telescope was pointed at disk centre
for more than 20~min. Each point in the plot corresponds to one
exposure.  Pointing errors should be negligible here. As seen from
the continuum contrast, the seeing was varying but improved during the
period. The line-centre variations show clear evidence of 5-minute
oscillations. The continuum intensity is clearly correlated with
these. The variations of the other line parameters do not seem to
correlate with oscillations but seem consistent with being caused by 
granulation -- mainly from its time evolution but probably with
contribution from slit drifts and rotation relative to the solar image.
There is a noise envelope, probably caused by a combination of photon
noise and seeing jitter, totally amounting to about 1\% in equivalent width.

\subsubsection{Summary}
All in all, the differential approach taken here should minimise many
systematic errors. The results from the tests are consistent with the
dominant errors being from solar variations and from pointing.
These are random in nature, and given the number of data sets, it seems reasonable
to use the spread of the individual data points -- graphically shown
in Fig.~\ref{fig:plotw} -- to estimate
the uncertainties in the mean values using the standard error of the mean
{\dk computed separately for the S+N data points and the W+E datapoints.}

  \begin{figure*}
   \centering
\hbox{
   \includegraphics[width=0.48\textwidth]{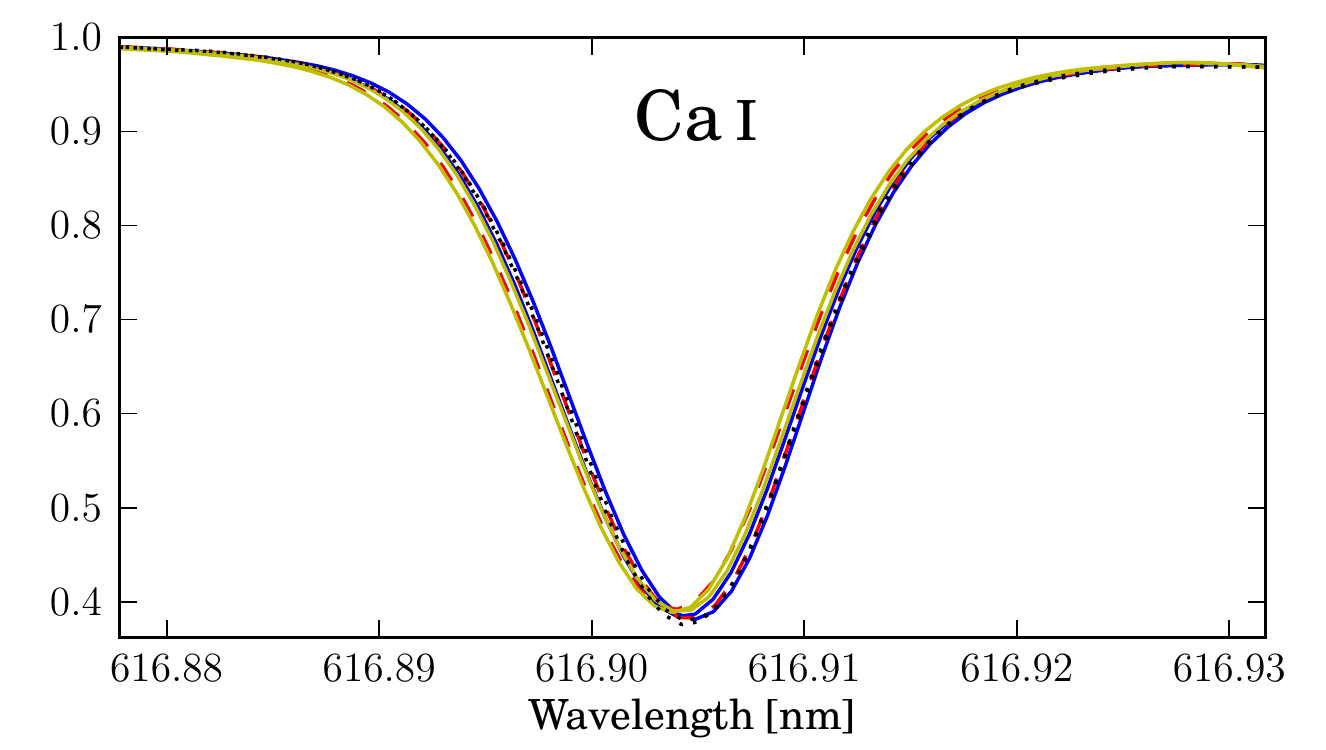}
   \includegraphics[width=0.48\textwidth]{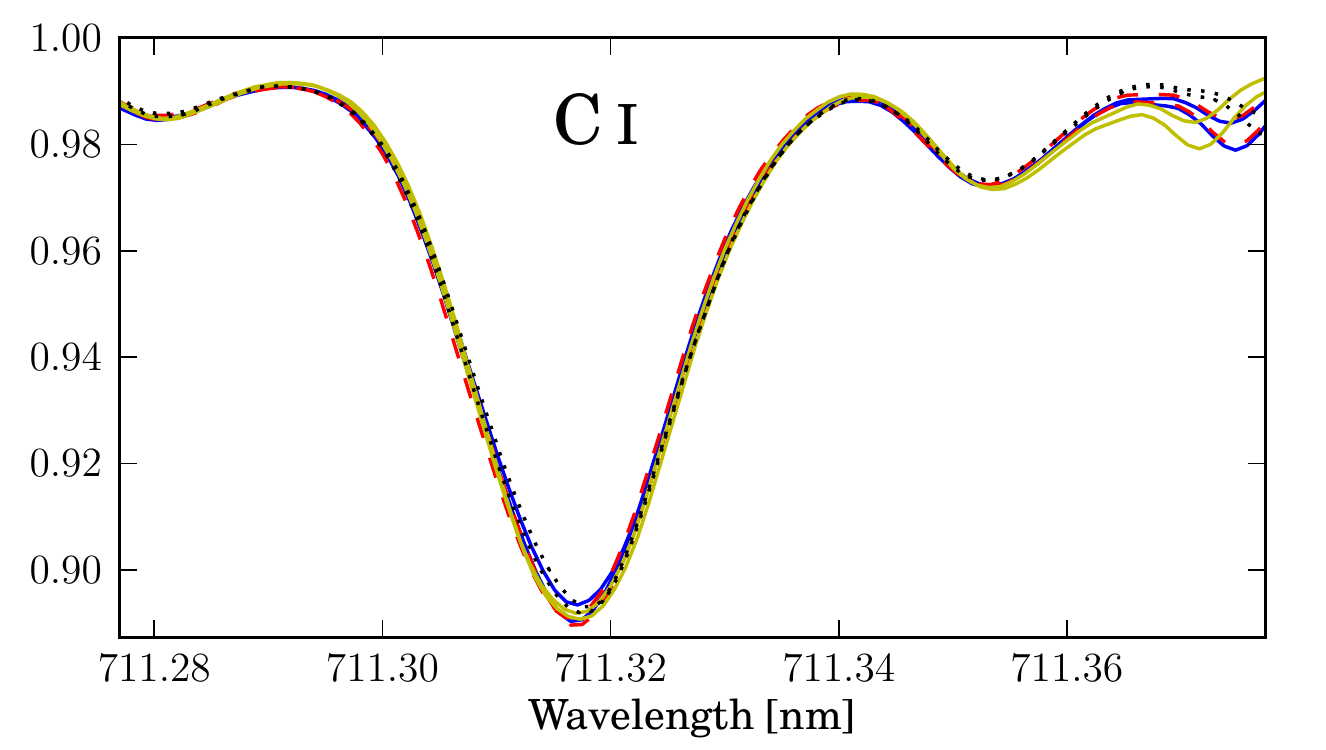}}
\hbox{   \includegraphics[width=0.48\textwidth]{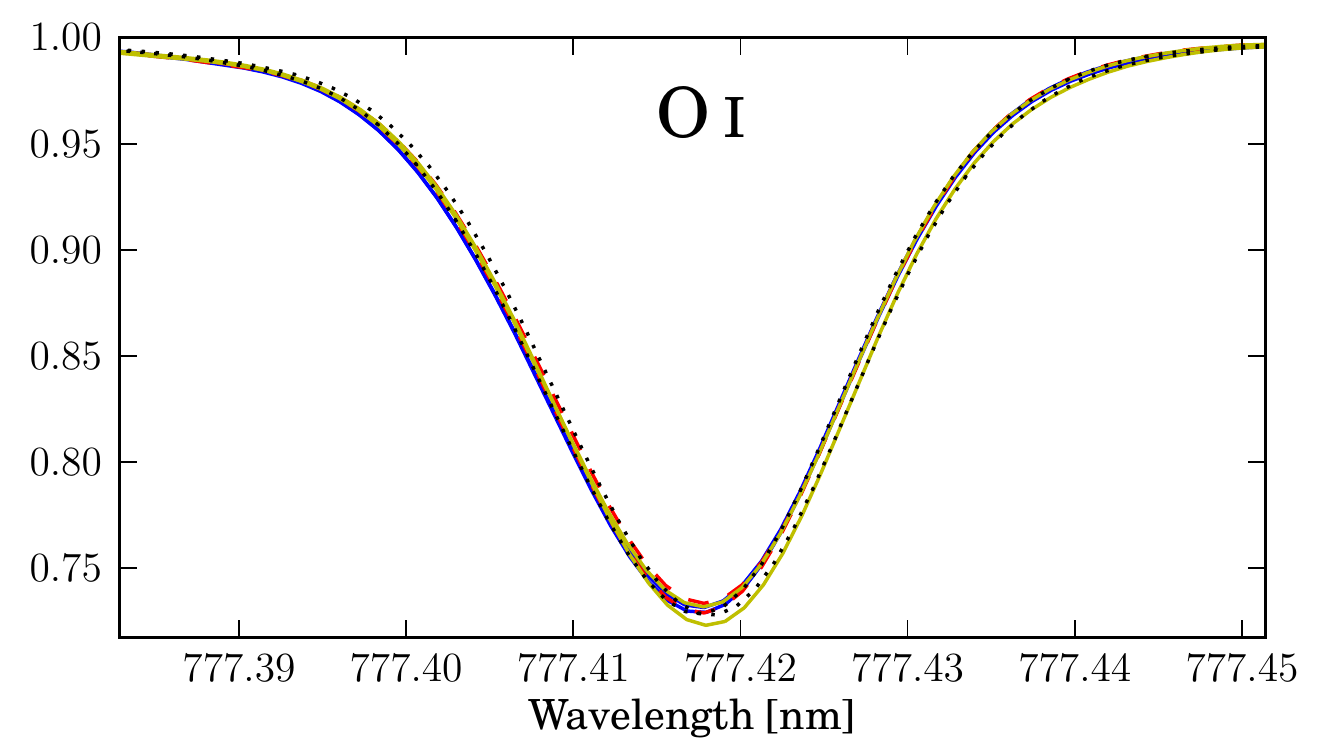}
   \includegraphics[width=0.48\textwidth]{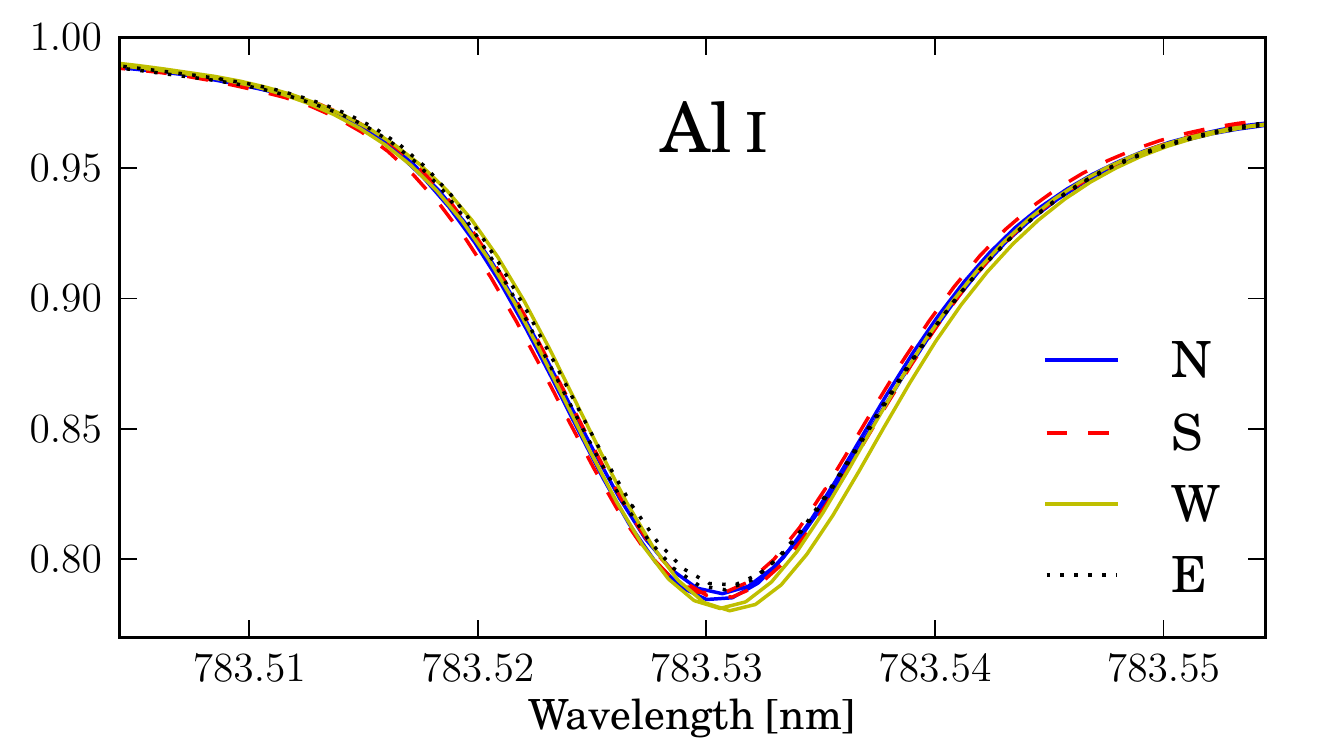}
}
   \caption{Two sample spectra for each of the four main disk positions acquired during the same day. The E and W spectra have been adjusted for solar rotation by applying a nominal Doppler shift of $\pm 1.89 {\rm km/s} \cdot \sin{45 \degr}$.}

              \label{fig:lat_var}%
    \end{figure*}


\section{Quiet Sun results}

\label{sec:qsresults}

   \begin{figure*}
   \centering
   \includegraphics[bb= 36 0 511 265]{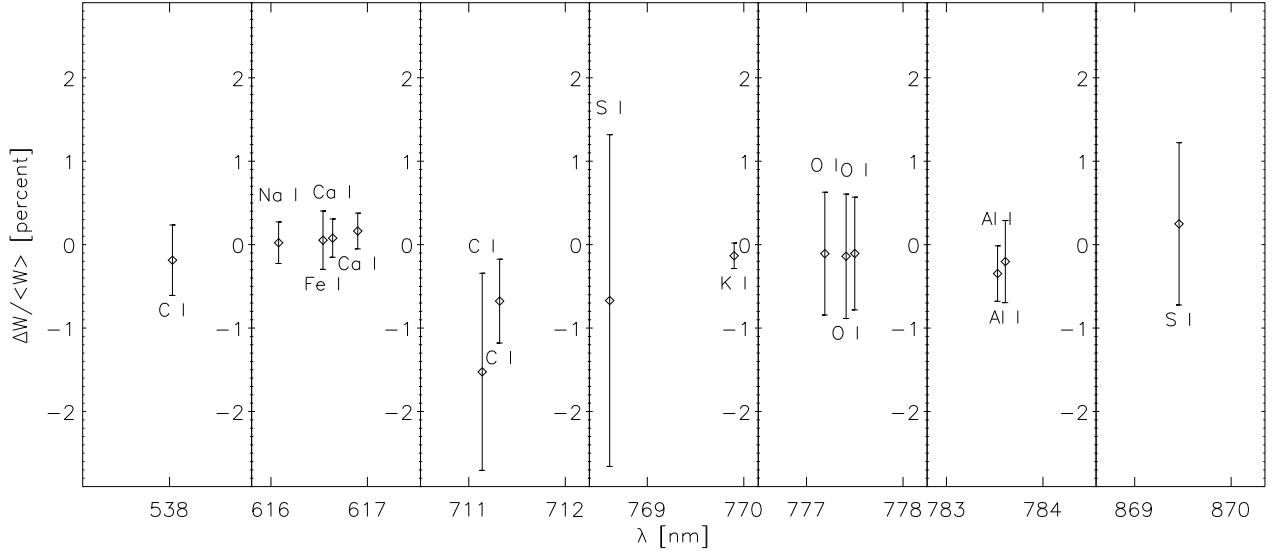}
   \caption{Difference between low-latitude and high-latitude
     equivalent widths for all lines. Error bars are based on the standard
     errors in the respective mean values.} 
              \label{fig:wdiff}%
    \end{figure*}

   \begin{figure}
   \centering
   \includegraphics[width=\columnwidth,bb= 0 0 270 215]{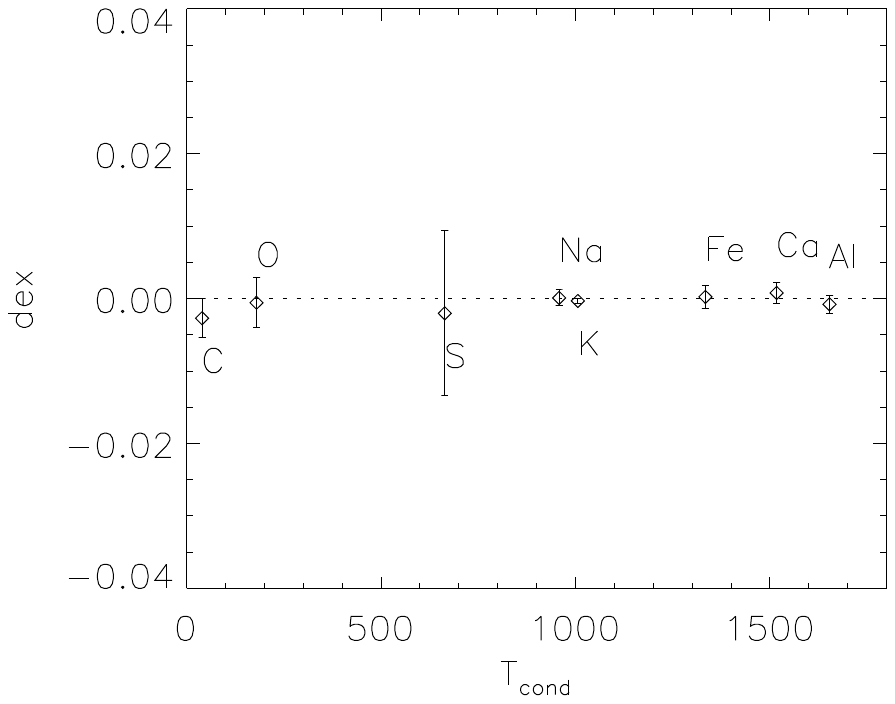}
   \caption{Difference between ``polar'' and ``equatorial'' abundances
     for all elements as function of their condensation temperatures
     according to \citet{2003ApJ...591.1220L}.}
              \label{fig:adiff}%
    \end{figure}
 
Figure~\ref{fig:plotw} displays the equivalent widths from all data
sets. Different symbols are used for the four different disk
positions. From the plots alone, no clear difference between the
different disk position is obvious for any line, with one possible
exception. We note that for the three \ion{O}{i} lines, four of the S
points lie below all of the N points. The fact that the fifth point
does not indicates that this N-S divergence may be due to activity or erroneous pointing
since the four deviating data sets were recorded on a different day.

Figure~\ref{fig:lat_var} illustrates the data points 
by showing examples of line profiles 
for the different disk position for four lines as recorded during the
same day.

Computing the mean values separately for the S+N and E+W disk positions, and
using the standard error in each  mean value as error estimate,
gives the result in  Figure~\ref{fig:wdiff}. Here, the relative difference in
    low-latitude (average of E and W) and high-latitude (average of N and S) equivalent width, $${\Delta W\over
      \langle W\rangle} = {(W_{low} - W_{high})\over \langle W
      \rangle},$$ is plotted  for each line.  The
    error bars are the standard errors in the mean values (1
    $\sigma$). The departure from zero of the
    relative differences $\Delta W/\langle W\rangle$ must be
    considered insignificant.

Our initial goal was to estimate the difference between the spectral
analysis an observer close to the equatorial plane of the Sun and one
who sees the Sun pole on. The polar observer will see a flux spectrum
which is characterised by our high-latitude spectra. The equatorial --
or terrestrial -- observer will see a flux spectrum that is a mix of
low-latitude and high-latitude contributions. We approximate this mix
to be 1:1. Thus the effective relative difference in flux equivalent
width is  
$${\mathbf{ \Delta W^{eff } \over \langle W\rangle} = {1\over 2}{\Delta W\over  \langle W\rangle}} .$$




The final result is shown in Fig.~\ref{fig:adiff} where the
difference in derived abundances per element is plotted as a function
of condensation temperature. The average values used for this plot
were computed without any weights.

The difference in derived abundance is computed with the help of
theoretical LTE curves of growth {\resub using} 
$\mathbf{\Delta W^{eff}/ \langle  W\rangle}$ 
as input. The difference between ``polar'' and
``equatorial'' abundances is within 0.005~dex for all elements and to
within 0.002 dex for the refractory elements. The error bars are
computed directly from those in Fig.~\ref{fig:wdiff}. It is clear that
there is no significant difference and that, in particular, no trend
with condensation temperature is present.

\section{Solar activity}
\label{sec:activ}
The principal goal of this work was to look for possible
latitude-effects in the quiet Sun and efforts were made to avoid any
activity. The result was negative, but that leaves the possibility
that the Sun, or a star, could have significant inclination-dependent
spectra due to activity. The point is that even if two stars have the same
level and kind of magnetic activity as the Sun, the active regions
tend to keep to lower latitudes.
Alternatively, one could argue that the studies by
 \citet{2009ApJ...704L..66M} and \citet{2009A&A...508L..17R} were made when the
 Sun was in an unusual deep activity minimum. It the solar twins
 happened to have systematically higher levels of activity, perhaps
 their spectra would be different, regardless of inclination angle.

An investigation of this possibility is beyond the scope
of this paper. (During our observing run, some data were in fact
acquired from regions of moderate activity but the amount of data is not
enough to allow any conclusions.)  However, some tentative remarks can
be made based on the existing literature.



It seems likely that lines that are strongly affected by activity also would display a 
variation with the solar cycle.
Indeed, the long-term monitoring of solar spectral features of
\citet{2007ApJ...657.1137L} shows that some strong lines with
chromospheric contributions display variations in phase with the solar
cycle. But among weaker, clearly photospheric, lines only \ion{Mn}{i}
539.47~nm shows equivalent-width variations of 2~\% during the cycle.
The reason why this line is special is explained by
\citet{2009A&A...499..301V} as due to its strong hyperfine broadening
leading to a lack-of-broadening-in-granules effect and very high
contrast of magnetic structures in the line.  Otherwise, \ion{C}{i}
538~nm was found to be constant by \citet{2007ApJ...657.1137L} within 1.5\% 
($\Delta \log \epsilon < 0.006$\,dex)
with no signs of cycles or trends as
were other photospheric lines, e.g. the 777-nm \ion{O}{i} triplet whose
equivalent widths all lie within their noise envelope of 1\%.

So, while the influence of magnetic activity on lines used for abundance analysis is
well worth more investigations, it is unlikely to explain abundance patterns like those
found by \citet{2009ApJ...704L..66M} and \citet{2009A&A...508L..17R}. 

\section{Discussion and conclusions}
\label{sec:concl}

We draw the following conclusions:

 The TRIPPEL Spectrograph performs well and has a high stability 
suitable for comparative studies of the kind presented here.

 The method of differential comparison is 
  precise and insensitive to, e.g., seeing variations.

 The difference in equivalent widths for the lines observed at
  the equator and at latitudes around 45\degr is less than $1 \%$
 at
  $\mu = \sqrt{2}/2$, 
 {\dk except
  for the telluric-afflicted \ion{C}{i}~711.1~nm line}.  Our results exclude the
  possibility that the effect found by \citet{2009ApJ...704L..66M} in
  the comparison of the composition of the Sun with those of the solar
  twins, for refractories relative to volatiles and amounting to about
  $20\%$ in elemental abundance, could be due to the systematically
  different aspect angles when observing the Sun relative to the
  twins.

Finally, we note that while we have argued that
it is unlikely that solar activity could modify our qualitative
conclusions, the effects of magnetic activity on abundances derived
from high-resolution spectra need further investigation.
A natural follow-up study to the current work would be to
apply the same methods to explore in detail the
  differences between photospheric spectra in solar active regions, as
  compared with inactive regions.
 In the present age of high-precision
 spectroscopy, in particular in detailed differential work, it may be
 important to take the activity state into consideration when
 detailed abundance analysis is made.

\begin{acknowledgements}
The Swedish 1-m Solar Telescope is operated on the island of La Palma by the Institute for Solar Physics of the Royal Swedish Academy of Sciences in the Spanish Observatorio del Roque de los Muchachos of the Instituto de Astrof{\'i}sica de Canarias.
BG acknowledges support from the Swedish Research Council, VR. J.M. thanks support from FAPESP (2010/17510-3). K.L. was funded by the European Community's Human Potential Program through the European Solar Magnetism Network (contract HPRN-CT-2002-00313). 
\end{acknowledgements}

\bibliography{bib-strings,svst,ads,latitude}

\end{document}